\documentclass{article}

\usepackage{PRIMEarxiv}

\usepackage[utf8]{inputenc} % allow utf-8 input
\usepackage[T1]{fontenc}    % use 8-bit T1 fonts
\usepackage{hyperref}       % hyperlinks
\usepackage{url}            % simple URL typesetting
\usepackage{booktabs}       % professional-quality tables
\usepackage{amsfonts}       % blackboard math symbols
\usepackage{nicefrac}       % compact symbols for 1/2, etc.
\usepackage{microtype}      % microtypography
\usepackage{lipsum}
\usepackage{fancyhdr}       % header
\usepackage{graphicx}       % graphics
\graphicspath{{media/}}     % organize your images and other figures under media/ folder

\title{Broadband second harmonic generation using fixed-period periodically poled lithium niobate
}

\author{
  Yongjoo Kim  \\
  Department of Electronic and Electrical Engineering \\
  University College London, London, United Kingdom \\
  \texttt{yongjoo.kim@ucl.ac.uk} \\ 
   \And
   Yijia Cai  \\
  Department of Electronic and Electrical Engineering \\
  University College London, London, United Kingdom \\
  \texttt{yijia.cai.19@ucl.ac.uk} \\
   \And
  Zhixin Liu \\
  Department of Electronic and Electrical Engineering \\
  University College London, London, United Kingdom \\
  \texttt{zhixin.liu@ucl.ac.uk} \\
}

\begin{document}
\maketitle

\begin{abstract}
Periodically poled lithium niobate (PPLN) is a widely used nonlinear optical device for second harmonic generation (SHG). Despite its wide adoption in commercial systems, its bandwidth for SHG is fundamentally limited by the quasi-phase matching condition. This can be overcome by aperiodic or chirped PPLN structures; however, such devices are typically custom-fabricated and not readily available commercially. In this study, we investigate an alternative approach to achieving broadband SHG by using a standard PPLN crystal containing multiple fixed poling periods. Broadband operation is realized by angle tuning the crystal relative to input beam in free-space. The effect of angle tuning is examined over a range of incident angles, and an 1.6x enhancement in SHG bandwidth is demonstrated. These results suggest a practical and efficient strategy for broadband SHG using standard PPLN devices.

\end{abstract}

% keywords can be removed

\section{Introduction}
Broadband second harmonic generation (SHG) is a key enabling technology for a wide range of applications, including ultrafast spectroscopy\cite{Maiuri2020}, quantum light source \cite{Fang2025} and self-referenced frequency combs \cite{Jankowski2020, Mayer2015}. For this purpose, periodically poled lithium niobate (PPLN) has been widely adopted due to its large second-order nonlinearity, mature fabrication technology, and high power conversion efficiency enabled by quasi-phase matching (QPM). In standard PPLN devices, however, the poling period is fixed, which restricts the QPM condition to a narrow spectral range. As a result, the achievable SHG bandwidth is typically limited to only a few nanometers, constraining broadband frequency conversion.

To overcome this limitation, aperiodic \cite{Jing2017} and chirped \cite{Chen2014} PPLN devices have been proposed as promising solutions for broadband SHG, as the poling periods in those devices are engineered to support QPM over an extended spectral range. Nevertheless, such designs generally require advanced and customized fabrication processes, reducing their accessibility and availability. Consequently, these approaches are not yet widely supported by commercial vendors, which constrains their practical adoption in many experimental settings.  

In this work, we propose and demonstrate an alternative approach for achieving broadband SHG using an off-the-shelf, fixed-period PPLN device. The PPLN used in this study is a commercially available multi-period crystal, consisting of 9 fixed poling periods arranged laterally across the crystal. By changing the crystal angle with respect to the input beam, the beam traverses different poling periods as it propagates in the crystal. Using this method, we achieved a bandwidth of 17.56 nm centered at 780 nm - an 1.6x improvement compared to the case with no angle tuning - with an input optical frequency comb spanning 30.8 nm, without the need for custom-fabricated PPLN devices. This approach provides a simple, flexible, and readily accessible option for broadband using standard PPLN devices.

\section{Methods}
In this study, an optical frequency comb centered around 1555 nm was employed as a source, and the output beam from the source was focused onto a PPLN bulk crystal in free-space, generating second harmonic (SH) light at 780 nm. The SH light was then coupled into a single-mode fiber (SMF) and delivered to an optical spectrum analyzer for characterization. Measurements were performed for various PPLN tilt angles relative to the input beam to investigate the effect of angle tuning on the SHG bandwidth. 

Fig. 1(a) illustrates the experimental setup. A commercially available 2.5-GHz-repetition rate mode-locked laser (MENHIR-1550, Menhir Photonics AG) served as a frequency comb source. A liquid crystal on silicon-based programmable optical filter was used to compensate for dispersion accumulated along the optical path after the laser. The optical power was amplified to 1 W using an Erbium-doped fiber amplifier (EDFA). The EDFA output was directed to a free-space optics stage, where the beam was focused onto the PPLN crystal. The crystal was mounted on a two-dimensional (X, Y) linear translation stage to allow adjustment of the crystal position relative to the focal plane and on a rotation stage to change the crystal tilt angle. The SH light from the crystal was relayed by a pair of lenses and then coupled into an 780-nm SMF.

The free-space optics stage was designed to achieve an optimal beam waist inside the crystal. For the 1-mm-long PPLN used in this study, a theoretical result \cite{BoydandKleinmann} predicts an optimal focal waist of 12.6 $\mu$m. In practice, the beam waist was slightly increased to 15 $\mu$m to reduce the peak intensity and avoid exceeding the crystal damage threshold. As shown in Fig. 1(b), illumination optics were designed using OpticsStudio (ZEMAX) to ensure a focal spot size of 15 $\mu$m across the wavelength range of interest (1500 nm, 1550 nm, and 1600 nm). Similarly, the relaying optics for the SH light from 750 nm, 775 nm, and 800 nm were designed to maximize coupling efficiency into the SMF.  

\begin{figure}[h]
    \centering
    \includegraphics[width=1\linewidth]{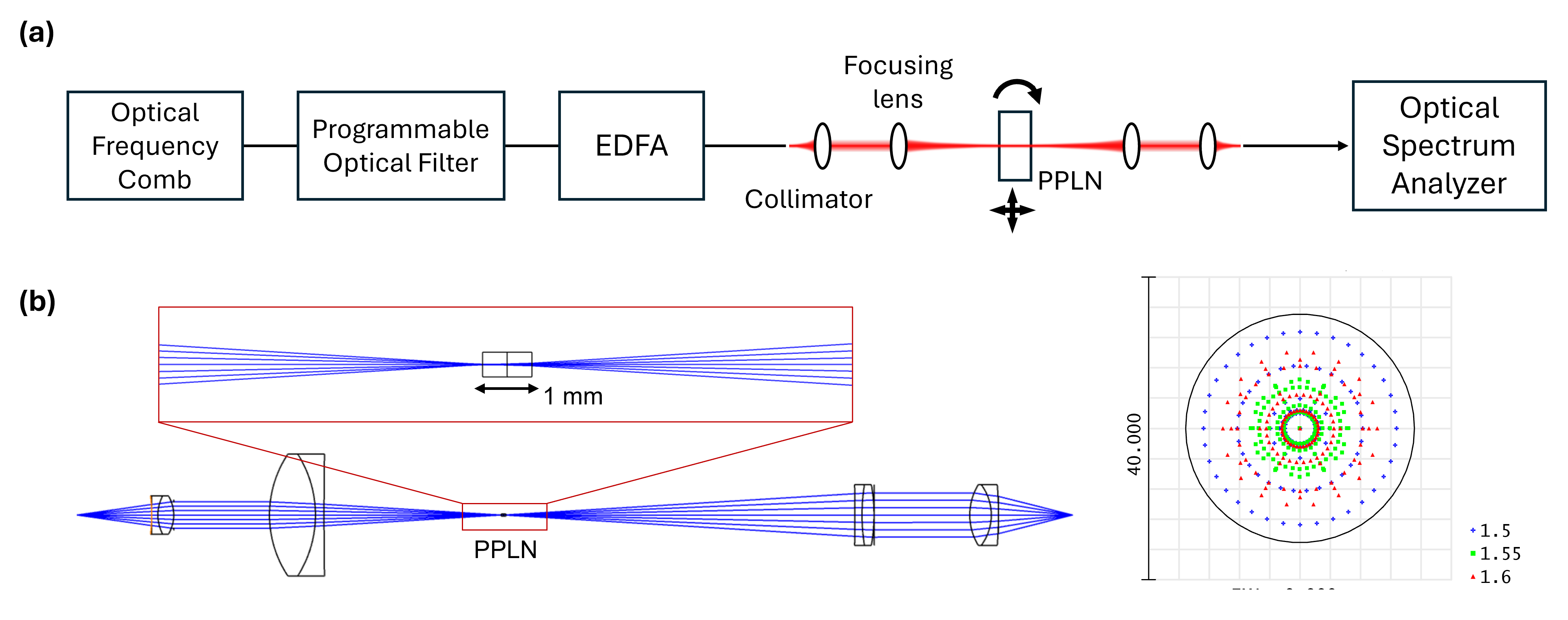}
    \caption{Experimental setup. (a) A schematic of setup. (b) (\textit{Left}) Optical layout of the free-space optics stage. \\  (\textit{Right}) Spot diagrams of 1500 nm (blue), 1550 nm (green), and 1600 nm (red) at the mid plane of the PPLN, confirming diffraction-limited performance of the optics design. The Airy radius (the black solid circle): 15.10 $\mu$m}
    \label{fig:placeholder}
\end{figure}

Fig. 2 depicts the geometric configuration of the optical beam relative to the PPLN crystal. The beam waist was positioned on a selected poling period that provided a central (SH) wavelength of 780 nm. Output spectra were then captured for tilt angle, $\theta$, from 0 to 40$^\circ$ at 10$^\circ$ increments. 

\begin{figure}
    \centering
    \includegraphics[width=0.5\linewidth]{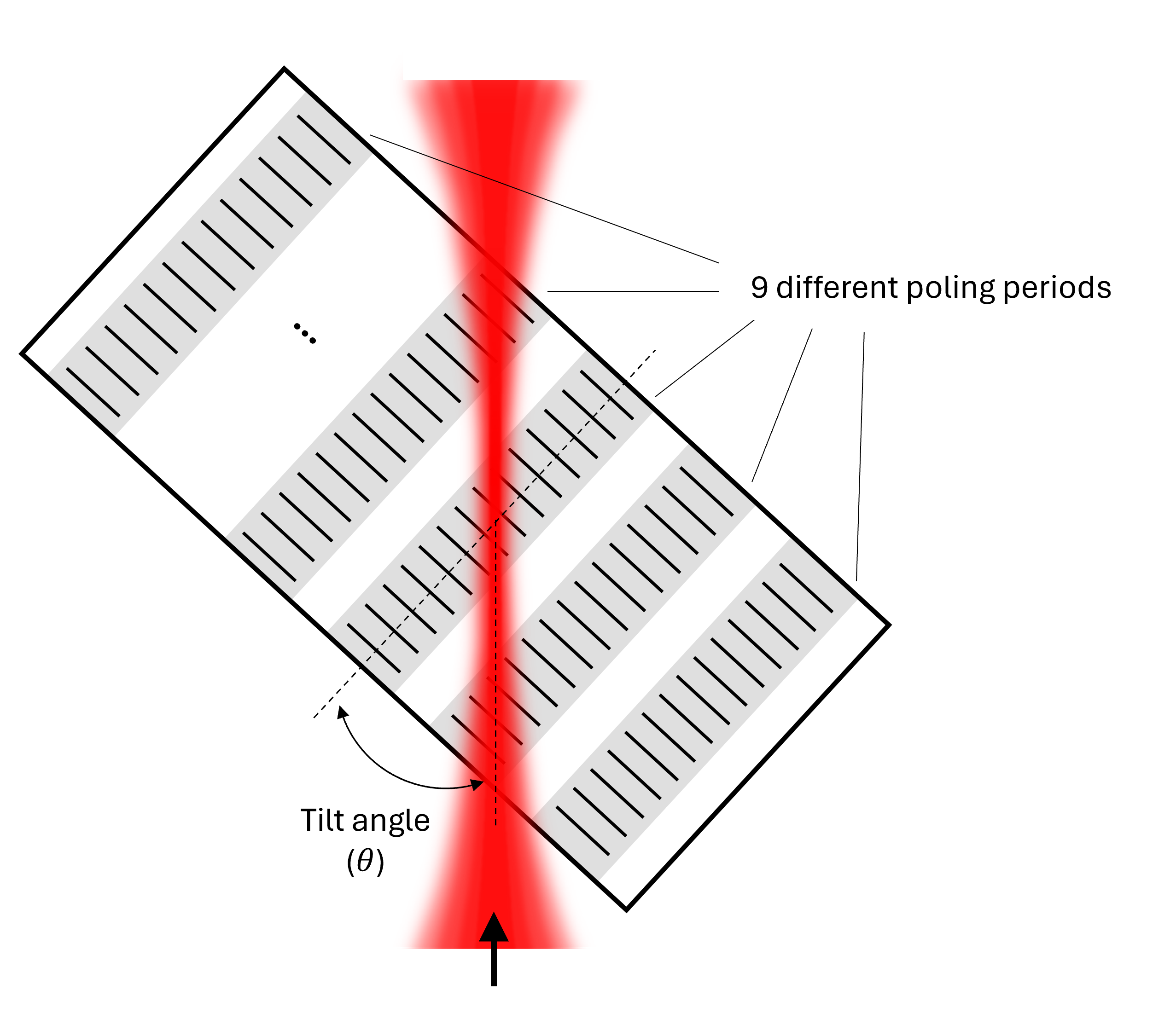}
    \caption{PPLN configuration for angle tuning}
    \label{fig:placeholder}
\end{figure}

\section{Results}
The EDFA output spectrum is shown in Fig. 3(a). The measured 10-dB bandwidth is 30.8 nm and is primarily limited by the EDFA gain bandwidth. Fig. 3(b) shows the SH output spectra . At $\theta$ = 0, the spectrum exhibits a main lobe around 780 nm with pronounced sidelobes. These sidelobes arise because the QPM bandwidth at this angle is not broad enough to accommodate the source bandwidth. The (10-dB) bandwidth of the main lobe is 10.66 nm. As $\theta$ increases to 10, 20, and 30$^\circ$, the sidelobes are suppressed effectively, resulting in broader bandwidths of 15.75 nm, 16.91 nm, and 17.56 nm, respectively. It is noted that the SH output power is inversely related to the bandwidth, as generally observed in the SHG process in PPLNs\cite{Chen2014}. We attribute this behavior to the combined effect of (1) a reduced effective interaction length between the beam and the grating within each individual poling period, which favors the SHG bandwidth at the expense of power conversion efficiency, and (2) coherent superposition of SHG contributions from successive grating periods along the beam path. Notably, the spectrum obtained at $\theta$ = 40$^\circ$ deviates from this trend. The underlying mechanisms for this observation will be further investigated. 

\begin{figure}[h]
    \centering
    \includegraphics[width=1\linewidth]{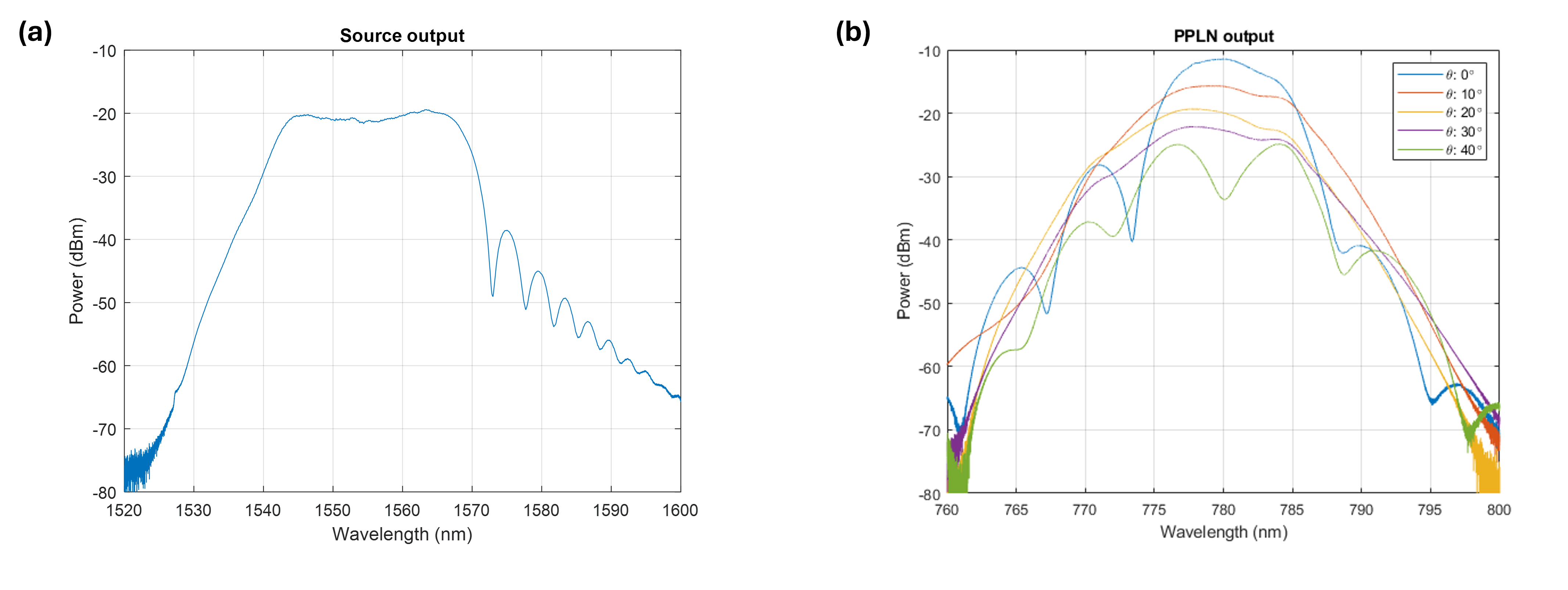}
    \caption{(a) EDFA output spectrum (b) PPLN output spectra for tilt angles from 0 to 40$^\circ$}
    \label{fig:placeholder}
\end{figure}

\section{Conclusion}
We demonstrated a SHG bandwidth of 17.56 nm centered at 780 nm using a multi, fixed-period PPLN pumped by an optical frequency comb centered at 1550 nm. This was achieved through angle tuning of the PPLN crystal, which enabled significantly broader SHG compared to the normal-incidence case. Broader SHG is expected with the use of a wider source spectrum, which remains future work. In addition, a more in-depth study - potentially including a theoretical modeling - of the impact of angle tuning on SHG performance in PPLN devices is needed. This will include the interaction of the beam waist,Rayleigh range, the spacing between each poling gratings, grating aperture sizes, and the crystal angle. Nevertheless, this work suggests that standard PPLN devices can support broadband frequency doubling, offering a simple and reliable solution for broadband frequency comb generation in the visible and near-infrared regions.

%%\section*{Acknowledgments}
%%This was was supported in part by......

%Bibliography
\bibliographystyle{unsrt}  
\bibliography{references}

\begin{thebibliography}{1}

\bibitem{Maiuri2020}
Margherita Maiuri, Marco Garavelli, and Giulio Cerullo.
\newblock Ultrafast spectroscopy: State of the art and open challenges.
\newblock {\em Journal of the American Chemical Society}, 2020.

\bibitem{Fang2025}
Xiao-Xu Fang, Guoliang Shentu, and He~Li.
\newblock Broadband quantum photon source in step-chirped periodically poled lithium niobate waveguide.
\newblock {\em arXiv}, 2025.

\bibitem{Jankowski2020}
Marc Jankowski, Carsten Langrock, Boris Desiatov, Alireza Marandi, Cheng Wang, Mian Zhang, Christopher~R. Phillips, Marko Lončar, and M.~M. Fejer.
\newblock Ultrabroadband nonlinear optics in nanophotonic periodically poled lithium niobate waveguides.
\newblock {\em Optica}, 2020.

\bibitem{Mayer2015}
A.~S. Mayer, A.~Klenner, A.~R. Johnson, K.~Luke, M.~R.~E. Lamont, Y.~Okawachi, M.~Lipson, A.~L. Gaeta, , and U.~Keller.
\newblock Frequency comb offset detection using supercontinuum generation in silicon nitride waveguides.
\newblock {\em Optics Express}, 2015.

\bibitem{Jing2017}
Jian Jiang, Jiandong Zhang, Kai Wang, Xuan Xiao, and Zuxing Zhang.
\newblock Broadband second-harmonic generation in appln with group-velocity matching.
\newblock {\em Optics Communications}, 2017.

\bibitem{Chen2014}
Bao-Qin Chen, Ming-Liang Ren, Rong-Juan Liu, Chao Zhang, Yan Sheng, Bo-Qin Ma, and Zhi-Yuan Li.
\newblock Simultaneous broadband generation of second and third harmonics from chirped nonlinear photonic crystals.
\newblock {\em Light Science \& Applications}, 2014.

\bibitem{BoydandKleinmann}
G.~D. Boyd and D.~A. Kleinman.
\newblock Parametric interaction of focused gaussian light beams.
\newblock {\em Journal of Applied Physics}, 1968.

\end{thebibliography}

\end{document}